\documentclass[12pt]{article}

\usepackage{epsfig}

\usepackage{amssymb}
\usepackage{amsfonts}

\usepackage{color}
\def\be{\begin{equation}}
\def\ee{\end{equation}}
\def\ba{\begin{array}{c}}
\def\ea{\end{array}}
\def\p{\partial}
\def\ben{$$}
\def\een{$$}

\newcommand{\bea}{\begin{eqnarray}}
\newcommand{\eea}{\end{eqnarray}}

\newcommand{\bbr}{\br\!\br}
\newcommand{\kkt}{\kt\!\kt}
\newcommand{\pbr}{\prec\!\!}
\newcommand{\pkt}{\!\!\succ\,\,}
\newcommand{\kt}{\rangle}
\newcommand{\br}{\langle}

\newtheorem{thm}{Theorem}

\newtheorem{lemma}[thm]{Lemma}

\newenvironment{proof}{\noindent
 {\bf Proof.}}{\hfill$\square$\vspace{3mm}\endtrivlist}

\begin{document}

\begin{center}

{\Large \bf

Quasi-Hermitian formulation of quantum mechanics
using two conjugate
Schr\"{o}dinger equations

}

\end{center}

\vspace{0.8cm}

\begin{center}

  {\bf Miloslav Znojil}$^{a,b}$

\end{center}

 $^{a}$
{Department of Physics, Faculty of
Science, University of Hradec Kr\'{a}lov\'{e}, Rokitansk\'{e}ho 62,
50003 Hradec Kr\'{a}lov\'{e},
 Czech Republic}

 $^{b}$
{The Czech Academy of Sciences,
 Nuclear Physics Institute,
 Hlavn\'{\i} 130,
250 68 \v{R}e\v{z}, Czech Republic,
{e-mail znojil@ujf.cas.cz}

 %$^{c}$
%{Institute of System Science, Durban University of Technology, Durban, South Africa}

\vspace{10mm}

%\newpage

%{}{...}

\subsection*{Abstract}

To the existing list of alternative
formulations of quantum mechanics
a new version of non-Hermitian interaction picture
is added.
What is new is that
{}{in contrast to the more
conventional non-Hermitian model-building recipes,
the primary
information about the observable phenomena
is provided not only
by the Hamiltonian but also
by an additional operator with real spectrum
(say, $R(t)$) representing another observable.
In the language of physics
the information carried by
$R(t)\neq R^\dagger(t)$ opens
the possibility of reaching the
exceptional-point degeneracy of the real eigenvalues,
i.e., a specific quantum phase transition.
In parallel,} the unitarity of the system remains guaranteed, as usual,
via a time-dependent
inner-product metric $\Theta(t)$.
{}{From the point of view of mathematics,
the control of evolution is provided
by} a pair of conjugate Schr\"{o}diner equations.
{}{This opens
the possibility
od an innovative}
dyadic representation of pure states {}{by which}
the direct use of $\Theta(t)$
is made redundant.
{}{The implementation
of the formalism is} illustrated via a schematic
cosmological toy model
in which the
canonical quantization leads to the necessity of working with
two conjugate
Wheeler-DeWitt equations.
From the point of view of physics, the
``kinematical input''
operator $R(t)$ may represent either the radius of
a homogeneous and isotropic expanding empty Universe
{}{or, if you wish, its Hubble radius, or
the
scale factor $ a(t)$ emerging
in the popular Lemaitre-Friedmann-Robertson-Walker classical solutions},  with the
exceptional-point
singularity of the spectrum {}{of  $R(t)$} mimicking the
birth of the Universe (``Big Bang'')
at $t=0$.

\subsection*{Keywords}

quantum theory of unitary systems;
non-Hermitian interaction representation;
non-stationary physical inner products;
dyadic representation of pure states;
schematic quantum model of Big Bang;

\newpage

\section{Introduction}

Around the turn of millennium it had widely been accepted that
the various existing
formulations of quantum mechanics (QM) ``differ dramatically in
mathematical and conceptual overview, yet each one makes identical
predictions for all experimental results'' \cite{Styer}. In the
cited review the authors emphasized the historical as well as
methodical importance of the Heisenberg's {\it alias\,}
``matrix'' formulation of QM (in which the states do not change in
time) as well as the economy of the most common
Schr\"{o}dinger's
{\it alias\,} ``wavefunction'' formulation (which ``shifts the focus
from measurable quantity to state'').

In {\it loc. cit.} the catalogue of formulations was not exhaustive.
The authors
did not mention the ``universal'' interaction picture (IP)
in which the observables (say, $\mathfrak{a}$) and the
states (say, $\psi$) are {\em both\,} allowed to vary with time,
$\mathfrak{a}^{(IP)}=\mathfrak{a}^{(IP)}(t)$
and $\psi^{(IP)}=\psi^{(IP)}(t)$.
In the general Hermitian IP framework of conventional textbooks \cite{Messiah}
one can easily re-derive both the Heisenberg-picture (HP)
or
Schr\"{o}dinger-picture (SP) methodical extremes
when setting $\psi^{(HP)}(t)=\psi^{(HP)}(0)=\psi^{(HP)}$
or $\mathfrak{a}^{(SP)}(t)=\mathfrak{a}^{(SP)}(0)=\mathfrak{a}^{(SP)}$,
respectively.

The review also did not reflect the
new quick developments in the field in the direction initiated
by Bender with Boettcher \cite{BB}.
The latter innovation turned attention to the
overrestrictive role played by the
Stone
theorem \cite{Stone}.
By this theorem, indeed,
the evolution described by Schr\"{o}dinger equation
 \be
 {\rm i}\frac{d}{dt}\, |\psi^{(SP)}(t)\pkt = \mathfrak{h}\, |\psi^{(SP)}(t)\pkt\,,
 \ \ \ \ \ \  |\psi^{(SP)}(t)\pkt \in {\cal L}^{(SP)}\,
 \label{1}
 \ee
is unitary in ${\cal L}^{(SP)}$ if and only if
the Hamiltonian is self-adjoint in ${\cal L}^{(SP)}$,
$\mathfrak{h}= \mathfrak{h}^\dagger$.
In our present paper we intend to offer a further extension
of the latter methodical developments
in which the Hermiticity restrictions imposed
by the
Stone
theorem
were circumvented.

In introduction we have to remind the readers
that the origin of the idea
can in fact be traced back to the paper by Dyson \cite{Dyson}.
Long before
the turn of millennium this author revealed that
the goal of having the theory non-Hermitian
but still unitary
can be achieved
via a non-unitary time-independent preconditioning
of the SP wavefunctions,
 \be
 |\psi^{(SP)}({t})\pkt\ \to \
  |\psi_{(Dyson)}({t})\kt = \Omega_{(Dyson)}^{-1} \,|\psi^{(SP)}({t})\pkt \
 \in \ {\cal H}_{(Dyson)}\,,
 \ \ \ \ \ \Omega_{(Dyson)}^{-1} \neq \Omega_{(Dyson)}^\dagger\,.
  \label{33}
 \ee
In applications, the non-unitarity
of the stationary Dyson's
map $\Omega_{(Dyson)}$ led to an efficient
description of correlations in various complicated
many-body systems \cite{Dyson,Jenssen,Geyer,Bishop}.

Along this line the potentially user-unfriendly
Hamiltonian $\mathfrak{h}$
has been replaced by its user-friendlier isospectral avatar
defined as acting in the new,
potentially user-friendlier Hilbert space ${\cal H}_{(Dyson)}$,
 \be
 \mathfrak{h} \to H_{}=
 \Omega^{-1}_{(Dyson)}\mathfrak{h}\,\Omega_{(Dyson)}\,.
 \label{parrs}
 \ee
The conventional Hermiticity
gets lost ($H\neq H^\dagger$)
so that Hilbert space ${\cal H}_{(Dyson)}$
has to be declared unphysical.
Importantly, {}{due to the time-independence of
$\Omega_{(Dyson)}$,}
the loss is just formal, with the Hermiticity
of $\mathfrak{h}$ in ${\cal L}^{(SP)}$ merely
replaced by the Dieudonn\'{e}'s \cite{Dieudonne} metric-mediated
quasi-Hermiticity of $H_{}$ in ${\cal H}_{(Dyson)}$,
 \be
 H^\dagger_{}\,\Theta_{(Dyson)}
 =\Theta_{(Dyson)}\,H_{}\,,
 \ \ \ \ \ \Theta_{(Dyson)}
 =\Omega_{(Dyson)}^\dagger\,\Omega_{(Dyson)} \neq I\,.
 \label{tretiqh}
 \ee
On a more abstract quantum-theoretical level the isospectrality
between a pre-selected, sufficiently user-friendly non-Hermitian
Hamiltonian $H\neq H^\dagger$ and its self-adjoint reconstructed
partner $\mathfrak{h}= \mathfrak{h}^\dagger$ opened multiple new
model-building strategies, first of all, in quantum field theory
\cite{BM,Carl}. The possibility of reconstruction of the ``missing''
physical inner-product metrics $\Theta$ from a given non-Hermitian
Hamiltonian $H$ led also, in the framework of relativistic QM, to a
completion of the years-long efforts of a consistent probabilistic
interpretation of the Klein-Gordon fields
\cite{KGali,KGalib,KGja,KGalic} and/or of the Proca fields
\cite{Smejkal,Smejkalb}.

A few other successful applications
can be found mentioned in the recent review of the field
by Mostafazadeh \cite{ali}.
Still, the author
had to admit there that
after a tentative transition from the Klein-Gordon equation
to a formally not too different Wheeler-DeWitt (WDW) equation
of quantum gravity \cite{Wheeler,DeWitt},
the applicability of the reconstruction of $\Theta=\Theta(H)$
appears to be limited. In the Mostafazadeh's own words,
``the lack
of a satisfactory solution of this problem has been
one of the major obstacles in transforming canonical
quantum gravity and
quantum cosmology into genuine physical theories''
(cf. p. 1291 in {\it loc. cit.}).

For this reason the review \cite{ali} of quasi-Hermitian QM
did only marginally mention the WDW models.
An analogous scepticism can be also found expressed
in the quantum-gravity-dedicated monographs \cite{Rovelli,Thiemann}).
The main mathematical obstacle can be seen in the fact that
the operators
``that arise in quantum cosmological models''
have to be manifestly time-dependent and that such a choice
``requires a more careful examination'' \cite{ali}.

More recently,
the problem has been reopened and
the latter challenge was re-addressed in
\cite{NIP}. Still,
the main methodical and conceptual challenges were,
from our present point of view,
circumvented.
For this reason we felt urged to complement the theory, in our present paper,
by a new analysis in which
the manifest
$t-$dependence of the quasi-Hermitian operators
would prove tractable in a more satisfactory manner.

{}{As we already indicated in Abstract,
an important source of the inspiration
of our present project was that
the vast majority of the
conventional applications of the
non-Hermitian model-building recipes
starts from the assumption of our knowledge of the
Hamiltonian $H$. In most cases, this operator is assumed observable, i.e.,
constrained by relation (\ref{tretiqh}).
At the same time, the more abstract theory of review \cite{Geyer}
admits the existence of an ``input information'' knowledge
of at least one other, independent
operator (i.e., in our present notation, of $R(t)$)
with real spectrum.}

{}{In some sense (cf. \cite{arabky}), an attempt of
a feasible and, at the same time, purposeful incorporation of
$R(t)$ in the formalism was one of the main driving forces behind
the present work. On the side of physics we decided to
motivate it by the needs of quantum cosmology in which
the notions like Hubble radius or scale factor
play a key role in the classical-physics
toy-model descriptions of
the empty, homogeneous and isotropically expanding Universe. Still,
for our present purposes we found it sufficient to speak just about
an entirely schematic observable ``radius of an expanding toy-model
Universe'' $r(t)$ which is allowed to vary with the so called
cosmological time $t$.}

The presentation of our results starts in
section \ref{Appendices} where we briefly review the
existing stationary and non-stationary versions
of the quasi-Hermitian quantum mechanics.
In section \ref{sedva} we then turn attention to the WDW equation and
review and
emphasize the recent progress in its study.
A deeper insight in its role is then provided in section \ref{orion}
in which we introduce our present highly schematic but instructive
toy model of the quantum Universe.

For the sake of simplicity, just the radius $r(t)$ will be
considered quantized, i.e., represented, just shortly
after Big Bang, by a
(quasi-Hermitian) operator $R(t)$. Subsequently, the related basic
technical questions of the construction of the physical
Hilbert-space metric $\Theta(t)$ and of the evolution equations in
the non-Hermitian interaction picture (NIP) are addressed and
reviewed in section \ref{Corio}. Several conceptual aspects of the
theory are finally discussed in section \ref{diskobol} and
summarized in section \ref{hlahol}.

\section{Two quasi-Hermitian formulations of quantum theory\label{Appendices}}

The introduction in quantum mechanics
usually starts, in textbooks, from its formulation in Schr\"{o}dinger
``representation'' {\it alias\,} ``picture'' (SP,  \cite{Messiah}).
In this language
the states are represented
by the ket-vector elements $|\psi^{(SP)}({t})\pkt$ of a suitable
Hilbert space ${\cal L}^{(SP)}$.
The unitary evolution of the system
is prescribed  by Eq.~(\ref{1}), i.e., by
Schr\"{o}dinger equation
in which the Hamiltonian is
required self-adjoint,
$\mathfrak{h}=\mathfrak{h}^\dagger$.
In such a setting
a decisive simplification of the solution of Eq.~(\ref{1})
can be achieved, in principle at least, via the (usually, just numerical)
diagonalization of the Hamiltonian.

\subsection{Non-Hermitian Schr\"{o}dinger picture (NSP)\label{AppendixA}}

In many realistic models the diagonalization of $\mathfrak{h}^{(SP)}$
may happen to be prohibitively
difficult.
More than half a century ago, fortunately,
Freeman Dyson
\cite{Dyson}
revealed that whenever
the ``maximal'' simplification
$\mathfrak{h} \to \mathfrak{h}_{(diagonal)}$ remains,
due to its complexity, unavailable,
the underlying realistic
Schr\"{o}dinger Eq.~(\ref{1}) might still be made tractable
via a ``partial'' simplification of the Hamiltonian. He just recommended that
the ``inaccessible'' diagonalization
is to be replaced by any other
(i.e., just invertible) auxiliary time-independent
isospectrality mapping $\Omega: \mathfrak{h} \to H$
which need not even be required unitary.

One just has to redefine the states
as well as, whenever necessary or useful, also the Hilbert space itself
(cf. Eq. (\ref{33}) above).
The original conventional Schr\"{o}dinger equation becomes replaced
by its equivalent representation in ${\cal H}_{(Dyson)}$,
 \be
 {\rm i}\,\frac{d}{dt}\,|\psi^{}(t)\kt
 =
 H\,|\psi^{}(t)\kt
 \,.
 %,
% \ \ \ \ H=\Omega^{-1}\,\mathfrak{h}\,\Omega
% \neq H^\dagger\,,\ \ \ H \neq H(t)\,.
% %\ \ \ \ H^\dagger\,\Theta=\Theta\,H\,.
 \label{hCauchy}
 \ee
As long as $H \neq H^\dagger$, it makes sense to
introduce
the ``alternative ket vectors''
 $$
 |\psi(t)\kkt=\Theta\,|\psi(t)\kt
 $$
evolving, in ${\cal H}_{(Dyson)}$, according to a complementary
Schr\"{o}dinger equation,
 \be
 {\rm i}\,\frac{d}{dt}\,|\psi^{}(t)\kkt
 =
 H^\dagger\,|\psi^{}(t)\kkt
 \,.
 \label{hCauchyb}
 \ee
The main benefit of the resulting formalism using two
Schr\"{o}dinger equations
and the two different Hamiltonians (viz., $H$ and $H^\dagger\neq H$)
it that for pure states,
the predictions of the results of measurements
have the fully analogous form in ${\cal L}$
and in ${\cal H}_{(Dyson)}$. Indeed,
once one considers any stationary and self-adjoint operator $\mathfrak{a}^{(SP)}$
representing an observable in ${\cal L}^{(SP)}$, and once one defines
its NSP avatar $A=\Omega^{-1}\,\mathfrak{a}^{(SP)}\,\Omega$
in ${\cal H}_{(Dyson)}$, the validity of
the elementary mathematical identity
 \be
 \pbr \psi^{(SP)}(t)|\mathfrak{a}^{(SP)}|\psi^{(SP)}(t)\pkt
 = \bbr \psi(t)|A|\psi(t)\kt
 \label{meas}
 \ee
implies the coincidence of the predictions (i.e., of the probability densities)
when computed
via the single textbook Schr\"{o}dinger Eq.~(\ref{1}) or via
the conjugate pair (\ref{hCauchy}) +
(\ref{hCauchyb}).

\subsection{Non-Hermitian interaction picture (NIP)\label{AppendixB}}

%\subsection*{A.1. Stationary limit:
%
%
%\subsection*{A.2. The full-fledged non-stationary NDP theory}

Once we admit the dependence of the Dyson's mapping on time,
$\Omega=\Omega(t)$,
it is sufficient to follow
the description of the necessary (i.e., NIP) amendments of the theory
as described in \cite{timedep,SIGMA}.
After such a generalization of the formalism the main changes
result from the emergence,
in both of the
non-Hermitian Schr\"{o}dinger equations,
of the non-vanishing Coriolis-force term
  \be
  \Sigma_{}(t)={\rm i}\,\Omega^{-1}(t)\,\dot{\Omega_{}}(t)\,,\ \ \ \
  \dot{\Omega_{}}(t)=
 \frac{d}{dt}\Omega_{}(t)
 \,.
 \label{aprek}
 \ee
The non-stationary generalization $H(t)$
of the observable non-Hermitian Hamiltonian
(or, in the standard language of mathematics, of the quasi-Hermitian
Hamiltonian) remains, by definition, isospectral with the self-adjoint
(though now, admissibly, non-stationary)
Hamiltonian $\mathfrak{h}(t)$ of textbooks.
Still, the price to pay for the
non-stationarity is that
the non-stationary upgrade of the two NIP Schr\"{o}dinger
equations reads
 \be
 {\rm i}\,\frac{d}{dt}\,|\psi^{}(t)\kt
 =
 G_{}(t)\,|\psi^{}(t)\kt
 \,,\ \ \ \ \ G(t)= H(t)-\Sigma(t)
 \label{Cauchy}
 \ee
 \be
 {\rm i}\,\frac{d}{dt}\,|\psi^{}(t)\kkt=
 G^\dagger_{}(t)\,|\psi^{}(t)\kkt\,,
 \ \ \ \ \ G^\dagger(t)= H^\dagger(t)-\Sigma^\dagger(t)
 \label{Cauchybe}
 \ee
where the generator of evolution
contains the Coriolis term $\Sigma(t)$
and ceases to be observable, therefore (cf., e.g., p. 1272 in \cite{ali}).

The admissibility of the non-stationarity of the prototype textbook
Hamiltonian $\mathfrak{h}(t)$ is, in some sense, exceptional
\cite{FrFrith}.
For
all of the other non-Hermitian
and non-stationary operators $A(t)$ of observables
defined, in ${\cal H}_{(Dyson)}$, by their
pull-down from ${\cal L}^{(SP)}$,
the formalism would become prohibitively complicated
unless we assume that
$\mathfrak{a}(t)=\mathfrak{a}(0)=\mathfrak{a}^{(SP)}$,
i.e., unless we require that
 \be
 A(t)=\Omega^{-1}(t)\,\mathfrak{a}^{(SP)}\,\Omega(t)
 \label{orefe}
 \ee
where the self-adjoint avatars
of the non-Hamiltonian observables
remain time-independent
(see also \cite{NIP} for a detailed discussion of this subtlety).

Under the latter assumption
one can
re-establish a complete parallelism between
the conventional Hermitian quantum
mechanics in interaction picture
and
its non-Hermitian alternative
in which
the two NIP Schr\"{o}dinger
Eqs.~(\ref{Cauchy})
and (\ref{Cauchybe})
control the evolution of states.
Naturally, a complete picture is only obtained when
one also takes into consideration the
manifest and necessary time-dependence of the observables.
The most straightforward guarantee of the
internal consistency of the theory is then provided by the
{}{following result.}

{}{\begin{lemma}
The time-dependence of observables (\ref{orefe})
can be reconstructed from their initial values
by the solution of
Heisenberg equation
 \be
 {\rm i\,}\frac{\partial}{\partial t} \,{A}_{}(t)=
  A(t)\,\Sigma(t) -\Sigma_{}(t)\,A(t)\,.
 %+K(t)\,,
% \ \ \ \ \ K(t)=\Omega_{}^{(-1)}(t)\,
% {\rm i\,}\dot{\mathfrak{q}}_{(SP)}(t)\,\Omega_{}(t)\,
 \label{beda}
 \ee
 \end{lemma}}
{}{\begin{proof}
Definition (\ref{orefe})
is equivalent to relation
$\Omega^{}(t)\,A(t)=\mathfrak{a}^{(SP)}\,\Omega(t)$
which is easily differentiated,
 $$
 \Omega_{}^{-1}(t)\,\frac{\partial}{\partial t} \,\Omega(t)
 +\frac{\partial}{\partial t} \, \,{A}_{}(t)
 =\Omega_{}^{-1}(t)\,\mathfrak{a}^{(SP)}\,
 \frac{\partial}{\partial t} \,\Omega(t)
 \,.
 $$
In the light of
definition (\ref{aprek})
of the Coriolis force this
immediately yields formula (\ref{beda}).
\end{proof}}

One has to notice that the role of the NIP generator of evolution
is played here by the Coriolis force.
The solution of the
equation specifies
the (by definition, non-stationary) operator $A(t)$
in consistent manner.
At the same time, our knowledge of this solution immediately
opens the possibility of an ultimate
evaluation of the
matrix elements entering the following
nonstationary upgrade of Eq.~(\ref{meas}),
 \be
  \bbr \psi(t_f)|A(t_f)|\psi(t_f)\kt
  =\pbr \psi^{(SP)}(t_f)|\mathfrak{a}^{(SP)}|\psi^{(SP)}(t_f)\pkt \,.
 \label{measb}
 \ee
This formula expresses the probabilistic contents of the
non-stationary theory and
quantifies the
prediction of the results of the measurement at time $t=t_f$.

\section{Samples of application\label{sedva}}

%The concept of background-independence: quantum gravity

In the preface to the ambitious theoretical
monograph \cite{Thiemann} we read that ``despite
an enormous effort of work by a vast amount of physicists over the
past 70 years, we still do not have a credible quantum general
relativity  theory'' (QGR).
``What we do have today are candidate
theories; \ldots for each of them one still has to show \ldots
that it reduces to the presently known \ldots classical general
relativity at low energies'' \cite{Thiemann}.

\subsection{Wheeler-DeWitt equation}

The incompleteness of the
candidate QGR theories
is best illustrated by the
canonical quantum gravity based on
WDW equation.
In this field,
incidentally, the progress is significant. On p. 1291 of \cite{ali},
for example,
it is emphasized that ``in the 1960's the discovery of the Hamiltonian
formulation of the General Theory of Relativity \ldots provided
the necessary means to apply Dirac's method of constrained quantization''.
We believe that in this context
also the NIP-based study of not too realistic
WDW equation did not still tell us its last word yet.

The scepticism of the theoreticians
is in a sharp contrast with
the {\em experimental\,}
side of the QGR problem
where the efforts of physicists were amazingly
successful. For example,
the
age of our Universe is currently widely agreed to be finite and
equal to cca 13.8 billion years \cite{Planck}.
It is worth adding that the
determination of the latter value belongs to one of the most impressive
recent experimental results in physics. Under a self-explanatory name
``Cosmic Background Radiation Anisotropy Satellite/Satellite
for Measurement of Background Anisotropies'' (COBRAS/SAMBA,
\cite{Planck})
the measurement was initiated around 1996 and
operated by the European Space Agency
between the years 2009 to 2013.
The necessary
sensitivity and resolution were further
improved by the NASA Wilkinson
Microwave Anisotropy Probe (WMAP).

This
resulted
in the data summarized in the so called
Lambda cold dark matter ($\Lambda$CDM) model
{\it alias\,}
``standard''
 cosmological model'' \cite{lam}.
In the acronym the first, Greek letter refers to the cosmological constant
while the use of the word ``standard''
emphasizes that its parameters fit not only
the expansion of the universe or
the distribution laws of the atomic nuclei and/or galaxies but also
the fairly contradictory hypothesis of
existence of the initial point-like Big-Bang
singularity.

Once we reopen the question of compatibility
of the $\Lambda$CDM hypotheses with
the basic principles of quantum theory we only have to return to
a moderate scepticism.
The applicability of the underlying classical-physics-based
concepts finds its first natural limitation in a restriction
to its far-from-Big-Bang verifications.
The experiments remain persuasively compatible with the classical
GR theory as the correct theory of gravity at macroscopic distances.

In this context we found our basic theoretical inspiration
and encouragement in
the comprehensive review paper \cite{ali}.
We read there that in quantum cosmology ``the
relevant \ldots second order differential equations''
(i.e., WDW equations)
resemble the Klein-Gordon equations and
``have the following general form''
 \be
 \frac{d^2}{dt^2}\, \psi^{}(t) +D(t)\,\psi^{}(t)=0\,
 \label{377}
 \ee
(cf. Eq. Nr. 377 and the related comments in \cite{ali}).
The symbol $\psi^{}(t)$ denotes here
a ``wave function of the Universe''
which would be ``void of a physical meaning''
without ``an appropriate inner product'' $\Theta$ \cite{ali}.

In the nearest future
the quantum effects emerging
at the singularities (as sampled by black holes or
hypothetical Big Bang)
have to be re-analyzed.
In other words, there is still a broad gap between our understanding
of the correspondence between the well-confirmed
classical singularities and
their internally consistent quantum analogues.
A model-based description
of their mechanism and dynamics
is still,  in the light of Eq.~(\ref{377}),
one of the most important subjects
of research and one of the sources of
open questions which motivated also our forthcoming considerations.

In a way explained in review \cite{ali}
(cf., in particular, section Nr. 9.2)
a full formal analogy between the Klein-Gordon and WDW
equations only exists when the operator part of Eq.(\ref{377})
remains time-independent, $D(t)=D(0)=D$.
In this case, indeed, the Klein-Gordon-type equation
can be transformed into its NSP equivalent
(\ref{hCauchy}) of section \ref{AppendixA}
(cf. also equation Nr. 378 in {\it loc. cit.}).
The second NSP
Schr\"{o}dinger Eq.~(\ref{hCauchyb}) of section \ref{AppendixA}
is then easily written in terms of the conjugate
Hamiltonian operator $H^\dagger$.
The correct physical (i.e., probabilistic)
interpretation of the evolution
then follows from the one-to-one correspondences
(\ref{33}) and (\ref{parrs}) between the states
and operators in the respective Hilbert spaces ${\cal L}^{(SP)}$ and
${\cal H}_{(Dyson)}$.

After a transition (of our present interest)
to the genuine WDW version of Eq.~(\ref{377})
the operator $D(t)$
must necessarily be kept manifestly time-dependent.
This forces us to make use of the
non-stationary NIP
formalism of section \ref{AppendixB}.
The main innovation
is that
{\em any\,} quasi-Hermitian
observable of interest (say, $A$ in Eq.~(\ref{meas})) becomes,
by definition, time-dependent (cf. Eq.~(\ref{orefe})).
As a consequence,
the prediction of {\em any\,} measurement
(i.e., the evaluation of the overlap~(\ref{measb}))
requires, suddenly,
not only the solution of the two
comparatively friendly
Schr\"{o}dinger-like
evolution equations for the state (i.e., the construction of the
two {\em vectors\,} in ${\cal H}_{(Dyson)}$)
but, first of all, also the solution $A(t)$ of another,
maximally unser-unfriendly Heisenberg-like
evolution equation (\ref{beda}) for the {\em operator}.

%\subsection{Hilbert-space problem}

\subsection{Closed versus open quantum systems}

Many authors
proved discouraged by the latter technical obstacles and
so they have
redirected their
attention, typically, to the
exactly solvable models
(later, we will pick up the letter \cite{FT} for illustration).
Frequently, people also simplify the model-building process by
giving up the unitarity.
They declare their quantum system,
in the spirit of Refs. \cite{Nimrodb,Nimrod}, ``open''.
In some sense, the newly acquired freedom
becomes abused because by definition of the open quantum systems
(formulated, basically, in the
Feshbach's spirit \cite{Feshbachu,Feshbach})
the resulting
``effective'' non-Hermitian
descriptions are, in the sense of fundamental theory,
incomplete \cite{Jones,Jonesb}.

{}{This being said one should add that
the use of the effective operators of observables really
enables one to pay more attention to the ever-present noise and
fluctuations in the quantum systems living in the real world
(cf., e.g., \cite{Spagnolo2,Spagnolo3}). In the future,
in this sense, it will certainly
be necessary to try to move beyond the restrictive
closed system models, indeed.}

{}{In our present paper} the consequent fundamental-theory
approach is not abandoned. {}{In its framework,
nevertheless,}
even the
operator-evolution
nature of the Heisenberg-like Eq.~(\ref{beda})
need not be the main technical problem.
Indeed, difficulties also emerge in connection with
the corresponding non-stationary
upgraded
doublet of Schr\"{o}dinger-like equations (\ref{Cauchy}) and (\ref{Cauchybe}).
As long as
the generator
$G(t)$
of the evolution of the states
becomes defined as the difference
between the Hamiltonian $H(t)$ (which is, by definition, observable)
and the Coriolis force $\Sigma(t)$ (which is, in general, not observable),
also the generator
$G(t)$ itself is not observable (once more, we may recall
Theorem Nr. 2 in \cite{ali} for details).
Moreover,
many examples (cf., e.g., \cite{2by2,3by3}) show that
the elements of its spectrum
need not even form the complex conjugate pairs.
For this reason, it makes also hardly any sense to try to
simplify the model by
imposing,
upon this operator, the popular ${\cal PT}-$symmetry constraint.
Still, in the search for innovations this is what is
often being done -- see,
e.g., \cite{FringF,PTFR}.

Recently, new light has been thrown upon these problems
by the two studies of several solvable and manifestly time-dependent
``wrong-sign'' anharmonic
oscillators \cite{FT,WS}.
Surprisingly enough,
it has been shown there that
it may still make good sense to
impose the ${\cal PT}-$symmetry constraint directly
upon the observable Hamiltonian $H(t)$.
At the first sight, the motivation seems missing because
this operator only enters the pair of the NIP Schr\"{o}dinger equations
in combination with Coriolis force.
Nevertheless,
the toy-model studies reconfirmed that
the main advantage of using the
concept of ${\cal PT}-$symmetry
lies in its capability of a clear
separation of the unitary
dynamical regime
(in which the symmetry remains unbroken)
from its unphysical
non-unitary
complement (in which the ${\cal PT}-$symmetry
becomes spontaneously broken).

\subsection{Pure states in dyadic representation}

A clarification of the
slightly complicated NIP situation becomes  provided
when one imagines that the Schr\"{o}dinger's, the Heisenberg's
and the Dirac's intermediate {\it alias\,} interaction
pictures describe the same physics.
All of them
characterize the same evolution of a quantum system which is
initiated by the preparation of a pure state $\psi(t)$
at $t=t_i=0$ and which finally leads to the
prediction of the results of the measurement at $t=t_f>0$.
In the language of mathematics this means that once we complement
the Dyson map (\ref{33})
(or, more precisely, its non-stationary, time-dependent amendment)
by a complementary, alternative replacement
 \be
 |\psi^{(SP)}({t})\pkt\ \to \
  |\psi_{(Dyson)}({t})\kkt = \Omega_{(Dyson)}^\dagger(t) \,|\psi^{(SP)}({t})\pkt \
 \in \ {\cal H}_{(Dyson)}\,,
  \label{33bei}
 \ee
we may immediately deduce that $ |\psi_{(Dyson)}({t})\kkt=
\Theta_{(Dyson)}(t)\,|\psi_{(Dyson)}({t})\kt$.
Thus, whenever we decide to work
with the two {\em different\,}
state-vector elements $ |\psi_{(Dyson)}({t})\kkt$ and $ |\psi_{(Dyson)}({t})\kt$
of the Hilbert space ${\cal H}_{(Dyson)}$,
it appears sufficient to work just with the information about the metric
encoded in the double-ket vector $ |\psi_{(Dyson)}({t})\kkt$.

The latter trick leads,
in a way emphasized in review \cite{NIP}, to
an enormous simplification of the NIP formalism.
The statement may be also given a
more compact mathematical form in which the pure state $\psi(t)$
is represented, in ${\cal H}_{(Dyson)}$,
by the rank-one (i.e., dyadic) projector
 \be
 \pi_\psi(t)=|\psi^{(Dyson)}(t)\kt\,
 \frac{1}{\bbr \psi^{(Dyson)}(t)
 |\psi^{(Dyson)}(t)\kt}\,\bbr \psi^{(Dyson)}(t)|\,.
 \label{epro3}
 \ee
With (or even without) the conventional assumptions of the biorthonormality and
bicompleteness
  \be
 \bbr \psi^{(Dyson)}(t)
 |\phi^{(Dyson)}(t)\kt=\delta_{\psi \phi}\,,
 \ \ \ \
 \sum_\psi\,
 |\psi^{(Dyson)}(t)\kt
 \bbr \psi^{(Dyson)}(t)| = I\,
 \label{thrup}
 \ee
this enables us to re-derive the fundamental measurement-prediction
formula (\ref{measb}).

\section{Quantum gravity in a toy model\label{orion}}

{}{In the preceding context it is worth
adding that in spite of a certain resemblance of our
state-representing formulae
with the
Aharonov's and Vaidman's time-symmetric two-state-vector
formalism \cite{Aharonov,Aharonov2},
the parallels are purely formal because the present
approach remains safely
traditional and time-asymmetric (see also
a few other comments in section \ref{usedl}
below). This means that we just stay in the framework of}
a traditional non-relativistic quantum cosmology in which
the $t-$dependent state (and, say, the pure state) of the Universe
would have to be prepared
at a suitable time $t=t_{initial}$.
One can expect
that the state of the Universe evolves
and gets measured at another instant $t=t_{final}>t_{initial}$.
Now, our task is to explain how one might realize such an evolution scenario,
in principle at least,
when making the pure states represented by the rank-one projectors
(\ref{epro3}).

\subsection{Classical
singularities}

There are not too many results in which
one would really succeed in making the
space-time background of QGR quantized (i.e., represented by an operator)
and,
simultaneously, time-dependent.
In our present considerations we decided to emphasize, therefore, just a few
preselected
methodical aspects of quantum gravity,
with our attention paid, predominantly, to
the requirement of the background independence
of the theory in which
even the measurements of the distances in an
empty Universe
would be of a strictly quantum, probabilistic nature \cite{gride}.

We will
test our ideas
in
the non-covariant kinematical regime in which
the time $t$ will still be treated as a parameter \cite{Hilgevoord}. Moreover,
even the strictly quantum Universe will be assumed
simplified and existing just
in a very small vicinity of its classical Big-Bang singularity.
After such a specification of
the simplified dynamical regime
we will add several further, methodically motivated
reductions of the picture.

\begin{itemize}

\item
The classical space-time
geometry of the Universe
has to remain
``next to trivial''.
We will employ the
not exceedingly revolutionary kinematics
working with the
non-covariant
concept of
absolute time.
The quantum-theory-controlled evolution of the Universe
will be then assumed unitary,
i.e., unitary in the language of the
more or less conventional quantum mechanics of
the so called closed systems.

\item
In both the classical and quantum settings, the naively physical non-relativistic
parameter of time $t$ will be assumed positive and set equal to zero at Big Bang.
On the classical non-relativistic level also the
3D spatial coordinates
will be assumed time-dependent, {}{therefore,}
$x=x(t)$, $y=y(t)$ and $z=z(t)$.

\item
All this would lead to a still nontrivial
version of the background independence
because
the observable
values of the spatial nodes
$x=x(t)$, $y=y(t)$ and $z=z(t)$
(i.e., say, point-particle positions \cite{Hilgevoord})
have to be defined (i.e., prepared
and/or measured) as eigenvalues of
operators, {}{in principle at least}.

\item
The {}{last} three operators have to be self-adjoint in
a physical Hilbert space ${\cal H}_{phys}$ in which the inner
product has the property of being time-dependent and degenerating at
$t=0$. In other words, a ``non-Hermitian'' NIP version of QM will
have to be used.

\end{itemize}

 \noindent
For the sake of simplicity of our toy-model-based considerations we
will assume that the kinematics of the expanding Universe
{}{will be just one-parametric. The purpose of such a choice
is twofold. In the context of mathematics a maximal simplicity of
our methodical considerations has to be achieved. In this sense one can simply
speak about
a homogeneous and isotropic, centrally symmetric
expanding empty Universe characterized, say,
by its volume or
radius $r(t)$.
Thus, in our present minimal project,
just such a real function would have to be
reinterpreted as an eigenvalue of an {\it ad hoc\,} operator $R(t)$.}

{}{In parallel, in the context of physics
the interpretation of the parameter $r(t)$ might be made more sophisticated,
with the details to be found, e.g., in the dedicated monograph \cite{Mukhanov}.
Thus, for example, one could identify its $t-$dependent value with the
scale factor $a(t)$ emerging
in the popular Lemaitre-Friedmann-Robertson-Walker classical solvable model,
or with the closely related function of $t$ called
Hubble radius, etc.}

{}{In any case,
the reduction of the description of the classical dynamics to a single
real parameter implies that in the centrally symmetric picture
with $r(t)=\sqrt{x^2(t)+y^2(t)+z^2(t)}$}
we will have to replace, firstly, the three spatial
Cartesian coordinates
$x=x(t)$, $y=y(t)$ and $z=z(t)$ by the equivalent
spherical coordinates $r=r(t)$, $\theta=\theta(t)$
and $\phi=\phi(t)$. Secondly,
for the sake of simplicity this will enable us to assume,
{}{in another reasonable approximation,}
the stationarity
$\theta(t)=\theta(0)$
and $\phi(t)=\phi(0)$ of {}{the angles}.
Moreover, we will
treat the latter two values fixed and not quantized.
Thus, both of the spherical angular coordinates
will be kept ``frozen'' and ``irrelevant'', i.e., classical and
time-independent.

{}{All of the
latter simplifications have just a methodical
motivation. In contrast,}
the radius of the Universe $r(t)$ itself will be
defined, after quantization, as one of the available real
eigenvalues of a time-dependent ``dynamical-geometry'' operator $R(t)$.
At any suitable Hilbert-space dimension
$N \leq \infty$ we will have to
write $r(t)=r_n(t)$ where the ``multiverse-counting'' quantum number
$n=1,2,\ldots, N$ specifies the
hypothetical ``prepared'' pure quantum state of the Universe.

The
quantum radius of the Universe $r_n(t)$ must be time-dependent. At Big Bang
we have to guarantee the existence of an ``unavoidable''  degeneracy
(also called ``exceptional point'', EP, \cite{Kato}),
$\lim_{t \to 0}r_n(t)=0$ for all $n$.
Our time-dependent ``dynamical-geometry'' operator $R(t)$ must be,
in our working Hilbert space ${\cal H}_{math}$,
non-Hermitian but Hermitizable {\it alias\,} quasi-Hermitian,
i.e., such that
 \be
 R^\dagger(t)\,\Theta(t)=\Theta(t)\,
 R(t)\,.
 \label{ablehe}
 \ee
In the literature,
interested readers may find a number of
the generic methodical comments on the latter equation
(cf., e.g., \cite{Geyer,ali,SIGMAdva}).
In what follows, we intend to work just with an illustrative
family of certain non-numerically tractable
$N$ by $N$ matrices $R(t)=R^{(N)}(t)$.
This will enable us to keep also the related discussion
sufficiently short and specific.

\subsection{The radius of the Universe in a solvable toy model\label{exCorio}}

A mathematical inspiration
of our present project of the realization
of a schematic quantum model of the Universe
dates back to {}{the unpublished} preprint \cite{1212.0734v2}.
We considered there, in an entirely different context,
a one-parametric family
of
non-Hermitian (but Hermitizable, quasi-Hermitian) matrices
with the real spectra
which represented the
discrete bound-state energies.
The purpose of the preprint
(to be cited as PI in what follows)
was a study of the slow, adiabatic unitary-evolution process
resulting in
a fall of the $N-$level systems into an
exceptional-point singularity (EPN, \cite{Kato}).

%with(linalg):
%>
%>tau:=1-t;
%> pu:=sqrt(3);
%>
%> rf:=sqrt(1-tau^2);ham:=matrix(4,4,
%>   [ -3+8*rf,pu*tau,0,0,
%>   -pu*tau,-1+8*rf,2*tau,0,
%>     0, -2*tau, 1+8*rf,pu*tau,
%>     0, 0,-pu*tau,3+8*rf
%>   ]);
%>gr:=eigenvals(ham);
%                  2 1/2            2 1/2             2 1/2
%  gr := 9 (2 t - t )   , 7 (2 t - t )   , 11 (2 t - t )   ,
%
%                  2 1/2
%        5 (2 t - t )
%
%> plot({gr[1],gr[2],gr[3],gr[4]},t=-.1..0.47,color=black,axes=framed);
%>
%
%********** Figure 1 zde
\begin{figure}[h]                     %instead of \begin{figure}[t]
\begin{center}                         %instead of \begin{center}
\epsfig{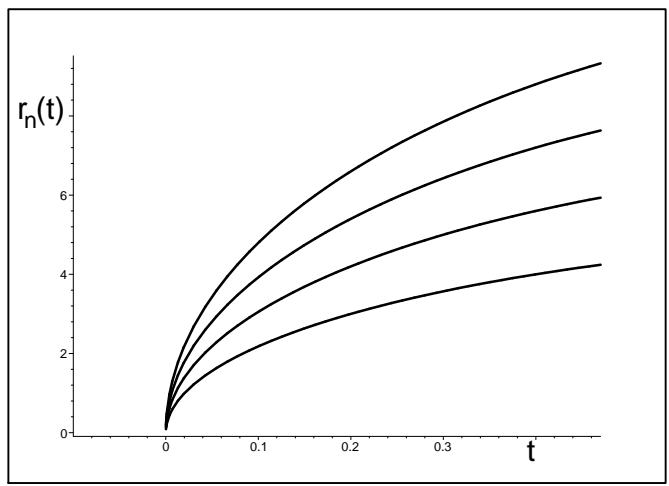}
\end{center}                         %instead of \end{center}
\vspace{-2mm}\caption{The ``multiverse'' eigenvalues
of the toy-model operator $R^{(4)}(t)$ of Eq.~(\ref{radaham})
representing the eligible instantaneous
size of the quantized Universe expanding after Big Bang.
 \label{figone}}
\end{figure}

Not the same but analogous matrices will be used
here in another role, viz., in the role
of a non-stationary operator $R(t)$
with the eigenvalues $r_n(t)$ representing the
observable instantaneous radii of the Universe.
For the sake of simplicity we will assume that
the dimension $N$ of
our schematic Hilbert space ${\cal H}_{(Dyson)}$
is finite. We will
consider $N=2,3,\ldots$ and
postulate that our
``kinematical'', geometric-background-representing input
matrices $R^{(N)}(t)$ have the following respective forms,
 \ben
 R^{(2)}({t}) = \left [\begin {array}{cc}
 -1+{\sigma^{(2)}(t)}&{\tau^{(2)}(t)}\\{}-{\tau^{(2)}(t)}
 &1+{\sigma^{(2)}(t)}\end {array}\right ]
 \,,\ \ \ \
 R^{(3)}({t}) = \left [\begin {array}{ccc}
  -2+{\sigma^{(3)}(t)}&\sqrt{2}\,{\tau^{(3)}(t)}  &0  \\
 -\sqrt{2}\,{\tau^{(3)}(t)}&{\sigma^{(3)}(t)}    &\sqrt{2}\,{\tau^{(3)}(t)}\\
  0&-\sqrt{2}\,{\tau^{(3)}(t)}&2+{\sigma^{(3)}(t)}
 \end {array}\right ]
 \,,\ \ \ \
 \een
 \be
 \ \ \ \ \
 R^{(4)}({t}) =
 \left [\begin {array}{cccc}
  -3+{\sigma^{(4)}(t)}&\sqrt{3}\,{\tau^{(4)}(t)}   &0  &0\\
 -\sqrt{3}\,{\tau^{(4)}(t)}&-1+{\sigma^{(4)}(t)}   &2\,{\tau^{(4)}(t)}  &0\\
  0&-2\,{\tau^{(4)}(t)}  &1+{\sigma^{(4)}(t)} &\sqrt{3}\,{\tau^{(4)}(t)}\\
  0&0&-\sqrt{3}\,{\tau^{(4)}(t)}&3+{\sigma^{(4)}(t)}
 \end {array}\right ]\,,\ \ldots\,.
 \label{radaham}
 \ee
Here, $\tau=\tau^{(N)}(t)$ and $\sigma=\sigma^{(N)}(t)$
are suitable real
and smooth functions of time $t$.
Thus,
for illustration we may choose the shift
$\sigma^{(N)}(t)=2N\,\sqrt{1-[\tau^{(N)}(t)]^2}$,
the dimension
$N=4$ and the parameter
$\tau^{(4)}(t)=1-t\,$. This would yield
the
spectrum $\{r_n(t)\}$
as displayed in Figure \ref{figone}.
We may see that the model is ``realistic'' in the sense that
at any quantum number $n$
our toy-model empty Universe exhibits a point-like singularity (Big Bang) at $t=0$
and a quick expansion at $t>0$.

A number of comments is to be made in advance. First,
we have to keep in mind that once we start from the
hypothetical knowledge of the kinematics,
it need not be easy
to combine the underlying space-evolution ansatzs
(i.e., in our toy-model case,
the specification of parameters in Eq.~(\ref{radaham}))
with the requirements of the dynamics
(sampled, in our case, by the WDW Eq.~(\ref{377})).

\section{The consistent model-building process\label{Corio}}

Our model-building
philosophy is based on the Big-Bang-admitting
ansatz (\ref{radaham}). Thus, our very first
task is to make the corresponding choice of the
kinematics (i.e., of the time-dependent matrix $R^{(N)}(t)$)
compatible with the unitarity of the evolution
(in this sense we assume that the quantum system under consideration
is a closed system).
This means that we have to take into account, first of all, the
Dieudonn\'{e}'s
Hermitizability constraint (\ref{ablehe}).

\subsection{The first step: The construction of the metric}

From the point of view of physics
all of the sufficiently ambitious quantum-Big-Bang-related  models
have to mimic
a quasi-static
phase transition.
{}{Naturally,
the concept of the
phase transition itself is very broad (see, e.g.,
the comprehensive review paper
\cite{Spagnolo}
where the authors list more than 400 further relevant references).}
In comparison, the range of physics behind our present
Big-Bang-related project is
{}{perceivably
narrower}. In a way discussed, more thoroughly, in
{}{
papers \cite{EPS,passage} we will
only deal here with the more specific
philosophy of the quantum phase transitions the realization of
which is based on the presence, in the space of parameters, of a suitable
Kato's \cite{Kato} exceptional point. Moreover,}
just a marginal attention will be
paid to the energy levels and to the Hamiltonians.
Our study will be redirected
to the
background-representing operator
(or rather non-Hermitian $N$-by-$N$-matrix)
$R^{(N)}(t)$. Depending on a real time-simulating parameter $t \in (0,1)$
(such that
$t=t^{(Big\,Bang)}=0$)
and preceding the case of a more realistic time-dependent
(i.e., non-relativistic)
{\em triplet\,} of $N$ by $N$ matrices
$X^{(N)}_1(t)$, $Y^{(N)}_2(t)$, $Z^{(N)}_3(t)$
representing a {\em dynamical, fully quantized\,}
(and, at finite $N< \infty$, just discretized)
three-dimensional space-time-grid background.

We will require that the
spectra of all of these matrices are complex at
the negative times $t<0$ (this
has to reflect the unobservable status of the space before Big Bang),
real but
EPN-degenerate at $t=0$ (i.e., at the hypothetical
non-relativistic Big-Bang instant)
and real and non-degenerate at $t > 0$
(for pragmatic reasons we will just keep in mind the not too large times,
i.e., say,
$t \leq 1$).
Moreover, in the
three-dimensional space
of the hypothetical expanding Universe
we will also reparametrize the coordinates and replace their time-dependent
and system-dependent
Cartesian grid $\{x(t),y(t),z(t)\}$ by the spherical triplet  $\{r(t),\theta(t),\phi(t)\}$.
For the sake of simplicity, the
angular coordinates will be assumed fixed,
and just the
radial one will be
treated as the expanding-Universe spatial background and
quantized,
i.e., treated as one of the eligible eigenvalues of an {\it ad hoc},
kinematical-input operator
$R^{(N)}(t)$.

For the methodical purposes (as well as for
the sake of definiteness) we will assume that the latter,
geometry-representing operator is given in advance, having the
form resembling closely the Hamiltonians in PI. Thus, once we
abbreviate $t=t(\tau)=1-\tau$ or introduce
a new variable $\tau=\tau(t)=1-t$,
the parallels become complete.

All of our toy-model matrices
will be chosen real, non-Hermitian and, whenever $t=1-\tau(t)>0$, Hermitizable. This means
that at any preselected matrix dimension $N$
there exists an inner-product-metric operator $\Theta=\Theta^{(N)}(t)$
(which need not be unique, see \cite{Geyer})
such that our spatial-grid-simulating operator $R=R^{(N)}(t)$
satisfies the quasi-Hermiticity condition
(\ref{ablehe}).
In PI we worked with the close analogues of our present matrices (\ref{radaham})
so that we can just recall and modify Theorem Nr 1 of PI and
formulate the following result.

\begin{thm}
At every finite Hilbert-space dimension $N<\infty$
the metric $\Theta^{(N)}({t})$ compatible with the respective
radii (\ref{radaham}) may be sought in the following generic form
 \be
 \Theta^{(N)}({t})=
 \sum_{j=1}^{N}\,{\cal M}^{(N)}(j)\,[-{\tau}(t)]^{j-1}
 \label{hlavni}
 \ee
containing the sparse-matrix coefficients
 \be
 {\cal M}^{(N)}(1)= \left[
 \begin {array}{cccc}
  \alpha_{11}(1)&0&\ldots&0
 \\
 {}
 0&\alpha_{12}(1)&\ddots&\vdots
 \\
 {}\vdots&\ddots&\ddots&0
 \\
 {}0&\ldots&0&\alpha_{1N}(1)
 \end {array} \right]\,,
 \label{hlavni1}
 \ee
 \be
 {\cal M}^{(N)}(2)= \left[
 \begin {array}{cccccc}
  0&\alpha_{11}(2)&0&\ldots&\ldots&0
 \\
 {}
 \alpha_{21}(2)&0&\alpha_{12}(2)&0&\ldots&0
 \\{}
 0&\alpha_{22}(2)&0&\alpha_{13}(2)&\ddots&\vdots
 \\{}
 \vdots&\ddots&\ddots&\ddots&\ddots&0
 \\{}
 0&\ldots&0&\alpha_{2,N-2}(2)&0&\alpha_{1,N-1}(2)
 \\
 0&\ldots&{}\ldots&0&\alpha_{2,N-1}(2)&0
 \end {array} \right]\,,
 \ \ \ \ \
 \label{hlavni2}
  \ee
 \be
  \ \ \ \ \
 {\cal M}^{(N)}(3)= \left[
 \begin {array}{ccccccc}
  0&0&\alpha_{11}(3)&0&\ldots&\ldots&0
 \\
 {}
 0&\alpha_{21}(3)&0&\alpha_{12}(3)&0&\ldots&0
 \\{}
 \alpha_{31}(3)&0&\alpha_{22}(3)&0&\alpha_{13}(3)&\ddots&\vdots
 \\{}
 0&\alpha_{32}(3)&\ddots&\ddots&\ddots&\ddots&0
 \\{}
 0&\ddots&\ddots&0&\alpha_{2,N-3}(3)&0&\alpha_{1,N-2}(3)
 \\{}
 \vdots&\ldots&0&\alpha_{3,N-3}(3)&0&\alpha_{2,N-2}(3)&0
 \\
 0&\ldots&{}\ldots&0&\alpha_{3,N-2}(3)&0&0
 \end {array} \right]\,,
 \label{hlavni3}
   \ee
etc.
\end{thm}

 \noindent
In PI we also recommended
to arrange the non-vanishing matrix elements into the $k$ by
$(N-k+1)$ arrays,
 \be
  \alpha{(k)}= \left[ \begin {array}{ccccc}
   \alpha_{11}(k)&\alpha_{12}(k)&\alpha_{13}(k)&\ldots&\alpha_{1,N-k+1}(k)
   \\
   \alpha_{21}(k)&\alpha_{22}(k)&\alpha_{23}(k)&\ldots&\alpha_{2,N-k+1}(k)
 \\
   \vdots& \vdots& \vdots& & \vdots
   \\
  % \alpha_{11}(k)&\alpha_{11}(k)&\alpha_{11}(k)&\alpha_{11}(k)
%   \\
  % \alpha_{k1}(k)&\alpha_{11}(k)&\alpha_{11}(k)&\alpha_{11}(k)&\alpha_{11}(k)
   \alpha_{k1}(k)&\alpha_{k2}(k)&\alpha_{k3}(k)&\ldots&\alpha_{k,N-k+1}(k)
 \end {array} \right]\,.
 \label{referendu}
 \ee
The values of these matrix elements had to be computed
as solutions of Eq.~(\ref{ablehe}) --
for the first few Hilbert-space dimensions $N$,
the results may be found in PI.

\subsection{Coriolis force and the evolution equations\label{reCorio}}

At a fixed Hilbert-space dimension $N$ the general inner-product-metric
solution of
Eq.~(\ref{ablehe})
is non-unique. It varies with an $N-$plet of
free parameters, the variability of which being only restricted by the condition
of the necessary positivity of the metric.
In \cite{passage} we studied
a special ``zero-spectral-shift'' sub-family of
our present time-dependent models (\ref{radaham})
with $\sigma^{(N)}(t)=0$.
At arbitrary $N$, we described there
certain ``optimal'' solutions (\ref{referendu}) which
exhibited a number of desirable features.
From our present point of view
the most important one was that the positivity of the metric
(i.e., of all of its  time-dependent eigenvalues $\theta_k^{(N)}({t})$)
proved guaranteed at all of the
dimensions $N>0$ and at all of the
times $t>0$ of interest.

Easily, the latter
result can be re-adapted to our present needs.
Irrespectively of the radius-positivity-guaranteeing {\it ad hoc\,}
shift parameters $\sigma^{(N)}(t)>0$,
the following Theorem can be easily proved by mathematical induction.

\begin{thm}
All of the time-dependent eigenvalues $\theta_k^{(N)}({t})$
of the optimal-radius-dependent inner-product metric
$\Theta^{(N)}(t)$ are given, at any matrix-dimension $N$, by
the following closed formula,
$$
\theta_k^{(N)}({t})=\sum_{m=1}^N\,C^{(N)}_{km}\,{[\tau(t)]}^{m-1}\,,\
\ \ \ \ \ \ k=1,2,\ldots,N
$$
where $ C_{1n}^{(N)}=\left (\ba N-1\\n-1 \ea \right )\,$, $
C_{2n}^{(N)}=\left (\ba N-2\\n-1 \ea \right )-\left (\ba N-2\\n-2
\ea \right )\, $ and, in general,
  $$
 C_{kn}^{(N)}=\sum_{p=1}^k\,(-1)^{p-1}\, \left (\ba k-1\\p-1 \ea
 \right )\, \left (\ba N-k\\n-p \ea \right )\,,\ \ \ \
 k,n=1,2,\ldots,N\,.
 $$

\end{thm}

 \noindent
The main consequence of this result is that all of the eigenvalues of the metric
are positive. This means that
we may
recall Eq.~(\ref{tretiqh})
(in which we reconstructed
the ``unknown'' metric $\Theta$
as a product of the two ``known'' Dyson maps)
and that we may try to invert the recipe (assuming that the metric
is ``known'' and that the Dyson map $\Omega(t)$
is to be reconstructed,
say, in the form of a real square root of $\Theta$ \cite{passage}).
This means that we can
factorize the metric
into the product
 \be
  \Theta_{}(t)
 =\Omega_{}^\dagger(t)\,\Omega_{}(t) \,
 \label{otretiqh}
 \ee
representing a
time-dependent generalization of the stationary factorization
formula of Eq. (\ref{tretiqh}).

This enables us to treat also the Coriolis-force matrix
(as defined by Eq. (\ref{aprek}))
as known. One can conclude that
the construction of the toy model is almost completed.
Indeed, any $J-$plet of its other observable features
can be represented by the respective operators (say, $\Lambda_j(t)$,
with, if needed, $j=0$ assigned to the energy-representing Hamiltonian).
All of these operators
must be, in terms of the same ``correct and physical'' Hilbert-space
metric, quasi-Hermitian,
 \be
 \Lambda_j^\dagger(t)\,\Theta(t) =\Theta(t)\,\Lambda_j(t)\,,\ \ \ \ j = 0, 1,
 \ldots, J\,.
 \label{quha}
 \ee
Secondly, any one of them (and,
in particular, the Hamiltonian $H(t)=\Lambda_0(t)$)
can be used to define the  basis which can be biorthonormalized
(cf. \cite{Brody} or Eq. (\ref{thrup})). The purpose may be served,
in the real-spectrum dynamical regime,
by the doublet of eigenvalue problems
 \be
 H_{{}}(t)\,|m^{{}}(t)\kt=E_m(t)\,|m^{{}}(t)\kt\,,
 \ \ \ \ \
 H^\dagger_{{}}(t)\,|m^{{}}(t)\kkt
 =E_m(t)\,|m^{{}}(t)\kkt\,,
 \ \ \ \ \ m=1,2,\ldots,N
 \,.
 \label{twoSE}
 \ee
Thirdly,
the knowledge of the metric also facilitates the
search for the other candidates for the observables
(denoted, say, as $\Lambda(t)$ without a subscript).
Indeed, once we consider the product
$\widetilde{\Lambda}(t)=\Theta(t)\,\Lambda(t)$,
we immediately see that $\widetilde{\Lambda}(t)=\widetilde{\Lambda}^\dagger(t)$
is Hermitian. Thus,
{\em any\,} Hermitian ``input-information'' matrix $\widetilde{\Lambda}(t)$
can be treated as a set of free parameters defining
a quasi-Hermitian operator
${\Lambda}(t)=\Theta^{-1}(t)\,\widetilde{\Lambda}(t)$
eligible as an observable.

The observables of the latter type
may be required to correspond to their conventional SP avatars
$\lambda^{(SP)}$ which are stationary, conserved and time-independent.
In such a case the process of the definition of the operator
(at all times) can be facilitated and replaced by the
definition of the operator just at a single instant $t=t_{initial}$,
with the completion of the construction of $\Lambda(t)$ (at all times)
provided by the solution of the corresponding Heisenberg Eq.~(\ref{beda}).

In the last step of our considerations
we may preselect a Hermitian matrix $A(t)$
and use it as the parameters defining the energy-representing
observable Hamiltonian
 $
 H_{}(t)=\Theta^{-1}_{}(t)\,A_{}(t)$. Then
we may immediately reconstruct the generator
$G_{}(t)=H_{}(t)-\Sigma_{}(t)$
of the evolution of the states which
enters, finally, the two conjugate Schr\"{o}dinger Eqs.~(\ref{Cauchy})
and (\ref{Cauchybe}).
The construction of the model is completed.

%\newpage

\section{Discussion\label{diskobol}}

In a way summarized in reviews \cite{Carl,ali,book}
the recent theoretical developments in quantum mechanics
threw new light on many traditional model-building strategies.
The main idea of the innovation
lies in an extension of the concept of the so called
observable
from its traditional form (i.e., from its self-adjoint representation) to an
unconventional alternative which is non-Hermitian but
which happens to be Hermitizable.
The Hermitization is still needed, mediated by an amended Hilbert-space
inner-product metric $\Theta$
which, ``it it exists'' \cite{Geyer}, varies with our choice of the
observable.

In applications one often works with
an observable Hamiltonian.
Whenever its most standard self-adjoint SP version
$\mathfrak{h}\neq \mathfrak{h}(t)$ happens to be
user-unfriendly, the desirable
user-friendliness can be recovered after transition to
its suitable non-Hermitian avatar.
A full compatibility of the resulting hiddenly Hermitian NSP
reformulation of quantum mechanics is achieved when, with a suitable
$\Theta=\Theta(H)$,
the new, non-Hermitian Hamiltonian $H \neq H^\dagger$  remains
$\Theta-$quasi-Hermitian,
 \be
 H^\dagger = \Theta\,H\,\Theta^{-1}\,.
 \ee
We already reminded the readers that
in section 9.2 of review \cite{ali}
it has been pointed out that the NSP-based construction of
the stationary inner-product matric
$\Theta=\Theta(H)$
plays a particularly important role in relativistic quantum mechanics
(with $H \neq H^\dagger$ being the Klein-Gordon operator)
and in the various versions of
application of the Dirac's method of constrained quantization
to gravity
(with $H \neq H^\dagger$ being the Wheeler-DeWitt operator).
It is only desirable to add now that
after a transition to the more advanced NIP version of the theory
in which one decides to work with the
time-dependent Hilbert-space metrics $\Theta(t)$,
most of the above-cited statements must be thoroughly reformulated.
In particular,
the most general non-stationary version of the Klein-Gordon operator
of the relativistic QM
cannot remain consistently
identified with the observable operator $H(t)$ anymore.

In our present paper the same change of
paradigm has been described and shown necessary also in the NIP approach
to the genuine, non-stationary Wheeler-DeWitt equation of quantum gravity.
For the purpose,
naturally, multiple technical simplifying assumptions had to be accepted.

\subsection{{}{Conventional
time-asymmetric} QM concept of the evolution\label{usedl}}

The description of a quantum system
in which the observables are represented by operators
is, certainly, richer than the description
of its classical limit \cite{Dorje}.
One of the related paradoxes
is that a substantial part of the success of
quantum theory
is, in some sense, serendipitious, based on a
lucky choice of one of many eligible ``quantizations''.
In this sense we are currently
not too lucky when trying to quantize
the Einstein's general relativity
(see, e.g., the Isham's foreword to the Thiemann's monograph
\cite{Thiemann}).

One of the problems is, in the Thiemann words,
that the ``quantum theory of the non-gravitational interactions
\ldots completely ignores General Relativity'' while the latter classical
theory {\it alias\,} geometry ``completely ignores quantum mechanics''
(see p. 9 in {\it loc. cit.}). In our present paper,
in this sense, we tried to stay, firmly,
in the framework of
non-relativistic quantum mechanics.

After such a simplification
the survival of the concept of time $t$
enables one to
order the evolution in a strictly causal manner.
Incidentally, the ``fixed-frame'' restriction
of such a type
can be softened
by a change of perspective
working with another, ``non-time'' evolution parameter \cite{gride}.
In an extreme case
as presented and discussed in methodical study \cite{Hilgevoord},
one can even
quantize the time itself,
i.e., one can treat $t$ as a
``pure-state'' eigenvalue of a ``quantum clock'' operator.

The idea of such a type
is presented also
in the Rovelli's monograph
\cite{Rovelli}. We can read there that
only in the conventional approaches one believes that
``the Schr\"{o}dinger picture is only viable for theories where there
is a global observable time variable $t$''. Naturally,
``this conflicts with GR [general relativity], where no such variable exists''
(cf. pp. 10 and 11
in {\it loc. cit.}).
One has to conclude that
a properly covariant formulation
of the unitary quantum evolution near Big Bang
is still not too well understood and formulated
at present, especially because
after the replacement of quantum mechanics by quantum field theory (QFT),
one reveals that also
``most of the conventional machinery
of perturbative QFT
is profoundly incompatible with the general-relativistic
framework'' \cite{Rovelli}.
{}{Thus, only the traditional,
perturbation-approximation-based
pragmatic approaches to the predictive cosmology seem
to be available at present
\cite{Mukhanov}.}

In this sense we proposed here that
one of the possible schematic keys to the puzzle
might be sought in the
quantization of the classical
GR singularities (like Big Bang) using, on quantum level,
the Kato's \cite{Kato} concept of the
exceptional-point degeneracy of the
schematic, non-covariant Universe at $t=0$.

\subsection{More realistic frameworks like loop quantum gravity}

The current progress in experimental astronomy
is amazing:
we already mentioned the
measurements of the cosmic microwave background
\cite{Planck}. This
confirmed the Big Bang hypothesis experimentally.
In parallel, its mathematically singular nature
also motivated an intensification of the efforts of making
the Einstein's classical general relativity (GR)
compatible with the first principles of quantum mechanics (QM)
\cite{Rovelli,Thiemann}.

The recent progress in this direction is remarkable.
We already mentioned
the studies of the conventional
canonical recipes aimed,
according to Wheeler \cite{Wheeler} and DeWitt \cite{DeWitt},
at the constructions of a ``wave-function of the Universe''.
Among the more recent related theoretical results
one must mention also the formalism of the so
called loop quantum gravity (LQG, \cite{Ashtekar}).
In this setting,
one is really able to work with the modified QM
called ``relational'', with some basic details mentioned
in section Nr. 5.6 of monograph \cite{Rovelli}.
Still, we read there that the relational
reformulation of QM ``has aspects that
need to be investigated further'' (cf. p. 367 in \cite{Rovelli}).

On these grounds our interest
in the problem has been born.
During one of the seminars on the subject
(dedicated to the description of
quantum Big Bang) we imagined that
people very often come to a quick
conclusion that the classical GR singularities
(like, typically, the Big-Bang-mediated
``abrupt'' birth of the Universe)
must {\it necessarily\,} get,
according to the conventional wisdom,
``smeared''
(i.e., in the mathematical sense, ``regularized'')
after quantization.

For a long time, the latter intuitive expectation had been widely
accepted.
A replacement of
the Big-Bang singularity
by the so called Big Bounce
 was
advocated
by the widest LQG community
\cite{Bounce,piech}.
Only very recently the assertion has been reconsidered and
opposed \cite{BBzpet}.
This means that the competition between the
Big Bang and Big Bounce hypotheses
may currently be considered reopened.

In our present toy model
the quantum
Big Bang instant remains singular.
Counterintuitive as such a possibility seems to be,
one could see its multiple analogues, say, in the
physics of phase transitions.
Naturally, many forms of the description of the
conventional phase transitions
are more or less standard, not
requiring the use of the
sophisticated mathematics of the LQG approach.
At the same time, the newly emerging undecided status of the
quantum Big Bang hypothesis
represents a challenge.
We believe that the new forms of insight were also provided
by our present paper.

\subsection{A broader physical context}

One of the main formal supports of optimism may be seen in the
fact that one of the key formal features of our present NIP theory
is in its richer representation of quantum dynamics.
Indeed, in the conventional version of QM
the flexibility of the model-building processes
is strongly restricted by the fact that
the (pure) state of a unitary quantum
system of interest is merely represented
by a ket-vector element $|\psi^{(SP)}\pkt$ of a preselected and
time-independent
Hilbert space
${\cal L}^{(SP)}$. In contrast,
the mathematical and phenomenological roles of the
ket-vectors in the NIP Hilbert
space ${\cal H}_{(Dyson)}$ become separated.
The amended theory works with the two non-equivalent
versions of the latter space, viz.,
with
${\cal H}_{math}$
(where the inner product in elementary but unphysical)
and with
${\cal H}_{phys}$.
In the latter case
one can say that either the definition of the
correct, physical inner product contains the operator
of metric, or that the operation of the physical Hermitian conjugation
is realized as the
less conventional antilinear map $|\psi\kt \ \to \ \bbr \psi|$.
This, indeed, simplifies the formalism
because the
mathematically user-friendly
space
${\cal H}_{math}$
(which must be declared ``unphysical'')
can also serve as a representation space for
${\cal H}_{phys}$.

From such a perspective the NIP approach comes with the new possibility
of making the family of the
gravity-related quantum field theories
``background-independent''(cf. p. 22 in \cite{Rovelli},
or the more detailed comments in \cite{Thiemann}).
From a purely pragmatic point of view this simply means that
in the conventional
models (i.e., say, in the point-particle wave functions $\psi(\vec{x})$)
even the parameters (i.e., in this case, the coordinates $\vec{x}$)
have to be perceived as eigenvalues of a suitable operator
(let us note that many of the associated
technical problems are discussed in the framework of the
so called non-commutative-geometry
\cite{Connes}).

In our present paper, an innovative realization of
the background-independence requirement has been achieved
by making the time-dependent radius of the expanding Universe quantized,
i.e., identified, in the pure-state multiverse-philosophy spirit, with one of the
eigenvalues of an {\it ad hoc\,} quasi-Hermitian operator $R(t)$.

%\newpage

\section{Conclusions\label{hlahol}}

At present, the use of non-Hermitian operators in quantum theory
is remarkably diversified, ranging from the traditional
and pragmatic effective-operator
descriptions of the open and resonant quantum systems  \cite{Nimrodb}
up to the new horizons opened by the studies of the
abstract mathematical aspects of the formalism \cite{book}.

In a narrower domain of the description of the closed (i.e., unitary)
quantum systems
using non-Hermitian operators
the main division line is the one which separates the
stationary and non-stationary theories.
In the former subdomain
the Coriolis forces
vanish so that
$H=G$.
There emerge no problems with calling
the Schr\"{o}dinger-equation generator
a Hamiltonian \cite{ali}.

In the latter, non-stationary-theory subdomain
the situation is different. We have to work there with
the less elementary relation
 \be
 H(t)=G(t)
+\Sigma(t)
\label{TDDE}
 \ee
(called, by some authors,
the time-dependent Dyson equation \cite{FringF,12} -- \cite{20}).
The term ``Hamiltonian'' must be then allocated, interpreted and used with
much more care \cite{NIP2}.

In the stationary NSP setting
the idea of acceptability of the various non-Hermitian
forms of quantum Hamiltonians
has its origin in the Dyson's
paper \cite{Dyson}.
The
knowledge of a standard stationary
self-adjoint Hamiltonian $\mathfrak{h}$ of textbooks
(which is, by definition, safely self-adjoint in ${\cal L}^{(SP)}$)
was simply complemented there
by a tentative, ``trial and error''
choice of $\Omega$. Via
the isospectrality constraint~(\ref{parrs})
one was immediately able to define a preconditioned,
friendlier stationary representation
$H$ of the conventional
Hamiltonian. This made the
innovative ``Dyson's picture'' of QM complete.

The encouraging experience with the
$\Omega-$mediated
simplifications of multiple conventional Schr\"{o}dinger equations
(say, in nuclear physics \cite{Jenssen}) inspired
Scholtz et al \cite{Geyer} to invert the paradigm. They
assumed that what we are given
are just the ``tractable'' time-independent operators of the observables
(including, first of all, the Hamiltonian
$H$) which are
non-Hermitian
but which possess the real spectra.
The core of the idea (i.e., of the
``quasi-Hermitian'' reformulation of quantum mechanics
called non-Hermitian Schr\"{o}dinger picture (NSP))
was that once we recall the respective quasi-Hermiticity
constraints (cf., e.g., Eqs.~(\ref{tretiqh}) or
(\ref{ablehe}) above), we may
reconstruct (not always uniquely)
and factorize (also not always uniquely) the correct
physical Hilbert-space metric $\Theta=\Omega^\dagger\,\Omega$
``if it exists'' (cf. p. 74 in \cite{Geyer}).
The resulting ``quasi-Hermitian-input''
version of the NSP formalism is then again a consistent theory.

The authors of paper \cite{Geyer}
were well aware of
the main weaknesses of their NSP recipe.
They identified them as lying,
in the sufficiently realistic models, not only in the ambiguity
of the assignment of $\Theta$ to a given Hamiltonian $H$
but also in the
technically rather complicated nature of
an explicit construction of
any such a metric
(cf. also a few related comments in \cite{ali}).
Fortunately,
a way out of the dead end has been found by
Bender with coauthors \cite{BB,Carl} who
proposed to narrow the class of the eligible
non-Hermitian stationary
Hamiltonians $H$.
The
more user-friendly
subfamily
of the
Hamiltonians was required
${\cal PT}-$symmetric, i.e., such that
 $
 H^\dagger {\cal PT}={\cal PT}\,H
 $.
Originally, the symbol ${\cal P}$ denoted here just the operator of parity
while the antilinear operator ${\cal T}$
mediated the time reversal.
Later, it became clear that after a suitable generalization
of these concepts, the
physics-motivated property of the
${\cal PT}-$symmetry of $H$
can be also perceived as mathematically
equivalent to the self-adjointness of $H$
with respect to a suitable
pseudo-metric, i.e., as the self-adjointness of $H$
in Krein space \cite{Langer,AKbook}.

The success of the ${\cal PT}-$symmetric models was enormous \cite{Carl}.
Paradoxically, it also
appeared to have the two not entirely pleasant consequences.
The first one was that around the year 2007 the mainstream research
left the
rather narrow area of
quantum physics.
Beyond this area
(i.e., typically, in classical optics) the idea of ${\cal PT}-$symmetry
found a large number of new and exciting
applications (for reviews see \cite{Christ} or the recent
monographs \cite{Christodoulides,Carlbook}).
The second paradox connected with
the deep appeal of
the idea of the ${\cal PT}-$symmetry of $H$
can be seen in the above-mentioned narrowing of the
scope and perspective.
In the words written on p. 1198 of review \cite{ali},
``the adopted terminology is rather unfortunate'' because
the ``${\cal PT}-$symmetric
QM is an example of a more general class of theories \ldots
in which ${\cal PT}-$symmetry does not play a basic role''.

As another unwanted consequence of the reduction of
the scope of the ${\cal PT}-$symmetric version of the theory
there emerged (and, for a long time, survived) several
``no-go'' theorems (sampled, e.g., by Theorem Nr. 2 in \cite{ali})
which claimed the impossibility of a sufficiently satisfactory
non-stationary
extension of the quasi-Hermitian quantum mechanics.
It took several years before the correct and consistent
non-stationary
extension of the quasi-Hermitian quantum mechanics
as described in \cite{timedep,SIGMA}
has finally been accepted as correct (cf., e.g., \cite{FringMou}).
The process of acceptance was also slowed down
by certain purely terminological misunderstandings
(cf., e.g., their brief account in \cite{NIP}).
At present, fortunately,
the situation seems clarified.
Different groups of authors
(using still very different notation conventions,
cf., e.g.,
papers \cite{Jub} or
\cite{PTFR})
accepted, ultimately, the same (or at least practically the same)
interpretation of the non-stationary NIP theory.

The related developments
enriched the field by a number of the new and highly relevant applications.
Virtually all of them can be
{}{characterized by}
the role played by
the time-dependent Dyson equation (\ref{TDDE})
{}{(cf., e.g., section Nr. 5 in \cite{WS})}.
The build-up of the theory may {}{then}
start either from the knowledge
of $H(t)$ (so that one can speak about a ``dynamical-information'' (DI) input),
or from the knowledge
of $\Sigma(t)$ (one then relies upon a purely kinematical
or ``Coriolis-force'' (CF) input information), or, finally,
from
$G(t)$ (let us call this option a ``Schr\"{o}dinger-generator'' (SG)
input knowledge).

In all of these alternative approaches their users decided
to call their preferred preselected component of Eq.~(\ref{TDDE})
``the Hamiltonian''.
In fact,
the above-cited words that
``the adopted terminology is rather unfortunate''
applied again.
The main reason is that even in the unitary evolution dynamical regime
the spectra of $\Sigma(t)$ and/or of
$G(t)$ need not be real
or even complex conjugate
in general \cite{2by2,3by3,NIP2}.
In this sense,
calling the generator $G(t)$ a Hamiltonian
(which was, originally, the proposal by one of my PhD students \cite{Bila,Bilab})
is far from optimal because only the spectrum of
the observable-energy component $H(t)$
of $G(t)=H(t)-\Sigma(t)$
can consistently be assumed real.

On these grounds
the most natural implementations of the NIP approach
seems to be provided by its DI model-building realization.
In
our recent paper \cite{WS}
such a conjecture has been tested using the exactly solvable
wrong-sign-oscillator model of
Fring and Tenney \cite{FT}.
We came to a not quite expected conclusion that
for the model in question,
by far the most convenient and efficient construction strategy
appeared to be the innocent-looking ``kinematical'' CF approach.

This observation can be perceived as one of the sources of inspiration
of our present paper. It forced us to
reconsider the theory and to
re-read one of the oldest studies in the field,
viz., paper \cite{Geyer} in which the authors always
kept in mind the need of
working with a complete
set of observables rather than just with a Hamiltonian.
We imagined that precisely this idea offers also
the
``missing source'' of a deeper understanding of the
non-stationary NIP
theory.

The return to the roots helped us to resolve at least some of the paradoxes.
For example, once one starts thinking about the unitary systems
characterized by more than one observable \cite{Geyer,arabky},
the build-up of the theory starting from the
mere single operator $H(t)$
appears to be {}{conceptually less satisfactory}.
During the build-up of a {}{more satisfactory} theory
one must keep in mind both
the dynamics (i.e., the influence of $H(t)$ upon the states $\psi(t)$
as mediated by Schr\"{o}dinger equation(s))
and
the kinematics ({}{due to the fact that}
$H(t)$ only appears in Schr\"{o}dinger equation(s)
in combination with Coriolis force).

In our present paper we managed to show that
the initial choice of a ``non-dynamical''
observable (i.e., in our present notation, of $R(t)$)
simplifies the constructions significantly.
This is, after all, our present main methodical
message.
We saw that our innovative strategy does not only simplify, decisively,
the ``introductory-step'' reconstruction of the kinematics
(i.e., of the metric
as well as of the Dyson map and of $\Sigma(t)$ from $R(t)$),
but that it also leaves
an entirely unrestricted space for
the subsequent choice of the ``dynamics'', i.e., for
an independent specification of the
instantaneous energy $H(t)$, etc.

We may only add that our other, serendipitious, physicists-addressing
message is that the independence of the initial choice of
the non-dynamical
observable
$R(t)$
might very well serve the purpose of the extension of
the applicability of the unitary NIP quantum theory
to the ``exotic'',
exceptional-point-related dynamical regimes. This is
sampled, in our schematic cosmological toy model,
by the demonstration of the possibility of an internal
consistence of the
hypothetical point-like Big Bang singularity
even after the quantization.

%\section*{Acknowledgements}

\end{document}